\documentclass{ws-mpla-arxiv} 
\usepackage{verbatim}
\begin{document}

\markboth{Gregor Herten}
{The First Year of the LHC}


\title{The First Year of the Large Hadron Collider: A Brief Review\footnote{Invited brief review, to be published in Mod. Phys. Lett. A, Vol. 26, No. 12 (2011) pp. 843-855.}}

\author{\footnotesize GREGOR HERTEN}

\address{Physics Department, Albert-Ludwigs-University Freiburg, Hermann-Herder-Str. 3 \\
79104 Freiburg, Germany\\
herten@uni-freiburg.de}

\maketitle


\begin{abstract}
The first year of LHC data taking provided an integrated luminosity of about $35~pb^{-1}$ in proton-proton collisions at $\sqrt{s}=7$~TeV. The accelerator and the experiments have demonstrated an excellent performance. The experiments have obtained important physics results in many areas, ranging from tests of the Standard Model to searches for new particles. Among other results the physics highlights have been the measurements of the W-, Z-boson and $t\bar{t}$ production cross-sections, improved limits on supersymmetric and other hypothetical particles and the observation  of jet-quenching, elliptical flow and $J/\psi$ suppression in lead-lead collisions at $\sqrt{s_{_{NN}}} = 2.76$~TeV.  

\end{abstract}


\section{History of LHC Operation} 	

After the formal construction approval of the Large Hadron Collider (LHC) by the CERN Council in its 100th session on December 16, 1994, more than 13 years were required to build the LHC\cite{LHC} and the four large experiments ALICE,\cite{AliceExperiment} ATLAS,\cite{AtlasExperiment} CMS,\cite{CmsExperiment} LHCb\cite{LhcbExperiment} as well as the smaller experiments LHCf\cite{LhcfExperiment} and TOTEM\cite{TotemExperiment}.  The injection of the first proton beams into the LHC was scheduled on September 10, 2008. Under close media coverage the LHC staff managed to inject and circulate single proton beams in both directions within a few hours. A few days later, on September 19, in an attempt to test super conducting LHC magnets at high currents, a quench occurred in sector 3-4,  most likely due to a faulty electrical contact between two magnets.  As a consequence, an electrical arc developed and punctured the helium enclosure, leading to release of helium into the insulation vacuum of the cryostat causing substantial damage to the magnets and to the beam pipe.

A subsequent investigation of the possible cause of this accident identified several critical areas, which needed repair and improvements before another attempt for proton-proton collisions could be started.  The repair and consolidation program consisted of  the replacement of 39 dipole magnets, 14 quadrupole magnets, cleaning of more than 4 km of vacuum beam tube and a substantial improvement of the quench protection system. Further studies revealed that most of the electrical contacts between magnets were of poor quality and needed repair before the magnets could be ramped to full current, corresponding to 7 TeV beam energy.  Since this repair would have required an additional one year of  shutdown, the CERN management decided to repair only the electrical contacts with the highest resistivities and operate the LHC for two years at a reduced beam energy of 3.5 TeV.

On November 20,  2009, a new attempt to inject proton beams started. Like in 2008, the LHC operators managed to accomplish their goals in a very short time. Single beams circulated in both directions within a few hours and the first collisions at injection energy (450 GeV) were recorded by the LHC experiments already after four days. Before the Christmas break, collision data sets were recorded at center-of-mass (cms) energies of 900 GeV and 2.36 TeV. The main data taking period in 2010 started with the first collisions at   a cms-energy of 7 TeV on March 30. The aim of the machine operators in 2010 was to increase the number of  protons per bunch, and successively throughout the year the number of bunches and the total stored energy.  Each increase of the stored energy was taken with great care to protect the experiments and machine components from beam loss. The specific luminosity increased from $10^{28} \mathrm{cm}^{-2}\mathrm{s}^{-1}$ in April to 
$\mathrm{2\times 10^{32} cm^{-2}s^{-1}}$ at the end of the proton run on November 4, exceeding the goal by a factor of two. The experiments recorded data with a total integrated luminosity of about 50 $\mathrm{pb}^{-1}$.  The prospects for the high luminosity run in 2011 are very promising, because the specific luminosity exceeded the goal and the emittance  is smaller than expected. 

After a fast switch from proton to ion operation, the first lead-lead collisions in the LHC at a nucleon-nucleon center of mass energy of 2.7 TeV were observed on November 8, 2010. In the following heavy-ion run until December 6, the experiments recorded data corresponding to an integrated luminosity of about 
$\mathrm{10 \mu b^{-1}}$.

\section{Performance of the Experiments}
 ATLAS and CMS are general purpose experiments with a nearly complete coverage in solid angle, allowing to reconstruct the transverse momenta of non-interacting particles, like neutrinos or stable neutralinos.  Typically the energy of hadronic particles  can be measured in a pseudorapidity range of $\vert\eta\vert < 5$, with $\eta\equiv - \ln \tan(\theta/2)$, where $\theta$ is the polar angle with respect to the beam axis. Charged leptons (electrons and muons) can be reconstructed up to $\vert\eta\vert < 2.5$.   The ALICE experiment is specialized on the study of heavy-ion collisions with excellent tracking and particle identification capabilities in a high multiplicity environment. LHCb is dedicated to the exploration of decays of hadrons containing $b$-quarks. Its main aim is the investigation of CP violation and rare $b$-hadron decays.   

Due to the forced LHC shutdown between September 2008  and November 2009, the experiments had ample time to gain experience in data acquisition, reconstruction and  analysis with extensive cosmic data taking. This period helped to identify  and repair detector problems, improve the data taking efficiency,  as well as the reconstruction software and detector simulation. Thus in November 2009, the experiments were in an unprecedented state of readiness, never seen before at the start of a new accelerator.  
  
The excellent performance of the experiments became clear after the first collisions were recorded. The trigger systems were able to cope with the increasing collisions rates throughout the year. All detector systems were able to record data with efficiencies larger than 90\%, most of them were close to 99\%.  The momentum and energy resolutions and particle identification capabilities improved steadily throughout 2010 as more and more data became available for calibration and detector studies. For all experiments the detector performance is very close to the expectation from Monte Carlo simulations. 

\begin{figure}[pht]
\centerline{\psfig{file=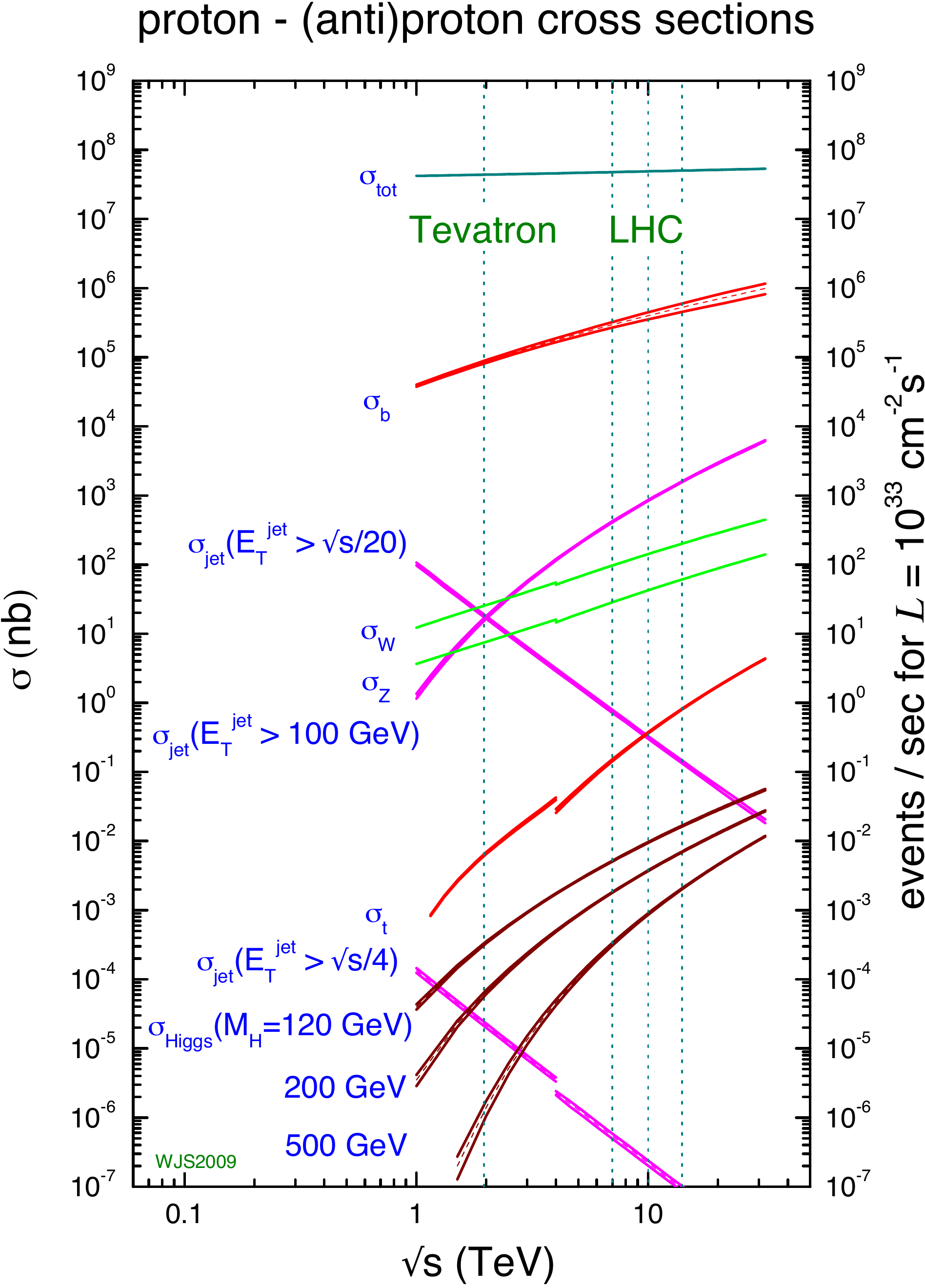,width=12cm,width=7cm}}
\vspace*{8pt}
\caption{Cross-sections for Standard Model processes at hadron colliders as function of the center-of-mass energy (W.J. Stirling, private communication).\protect\label{StirlingXsec}}
\end{figure}

\section{Physics Results}
The physics program in the first year of the LHC was guided by the cross-sections of various reference processes as show in Fig.~\ref{StirlingXsec}.  With increasing luminosity throughout 2010  the experiments have published results from high cross-section to lower cross-section processes, e.g. from multi-particle production (Òminimum biasÓ), inclusive particle spectra, $J/\psi$ and {\it Y}, W- and Z-bosons, top quark production and particle searches.  This review provides a selection of recent results from the LHC experiments. In total more than 100 results have been published based on the 2010 data set.  A complete list of publications from the LHC experiments can be found in Ref.~\refcite{LhcPub}. 

\subsection{Charged particle multiplicity} 
\begin{figure}[pht]
\centerline{\psfig{file=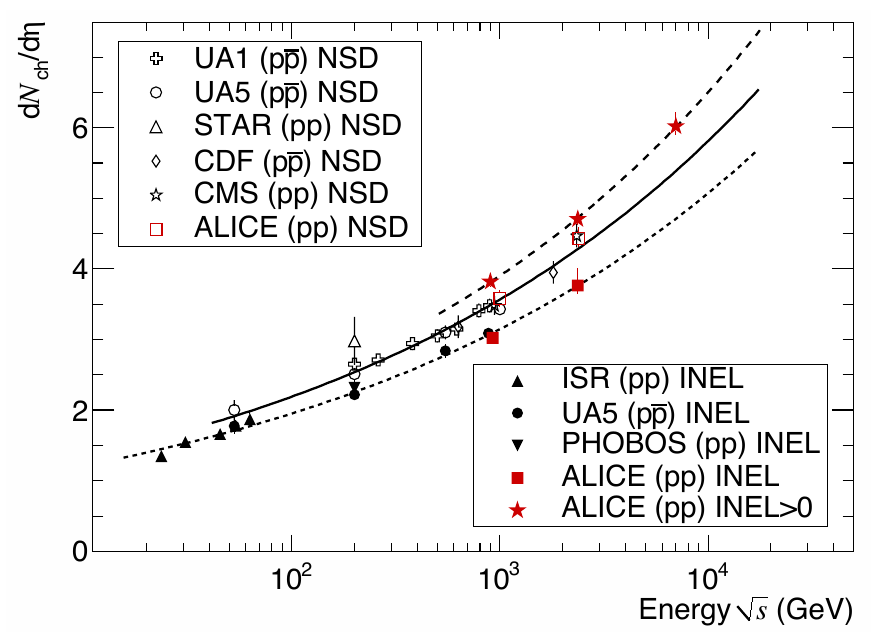,width=10cm}}
\vspace*{8pt}
\caption{ALICE measurement of the charged particle density at three center-of-mass energies.  For comparison with other experiments also the evolution for non-single diffractive (NSD) and inelastic events (INEL) is shown. The lines represent a fit through the measurements of NSD, INEL and INEL$>$0, respectively. 
\protect\label{AliceChargedParticles}}
\end{figure}
The first measurements of the LHC experiments were the study of multi-particle production, and the measurement of the charged particle multiplicities. These particle events are often called Òminimum biasÓ events because the trigger requires only a minimal energy deposition in the forward and backward region of the detector.  Such events  are modeled by inelastic proton-proton collisions and diffraction at small momentum transfer. Figure~\ref{AliceChargedParticles} shows the charged particle multiplicity as measured by the ALICE experiment at 0.9, 2.36 and 7 TeV from Refs.~\refcite{2010AliceChargedParticles1}-\refcite{2010AliceChargedParticles3}.  The increase in multiplicity from 900~GeV to 7~TeV is larger than predicted by presently used Monte Carlo models. The Pythia ATLAS-CSC tune is closest to the measurement. ALICE chose to normalize the measurements to an event class with at least one charged particle in the pseudorapidity range $\vert\eta\vert < 1$ in order to keep systematic effects due to the uncertainty of the contribution from diffraction small. A similar procedure has been published by the ATLAS collaboration over a wider pseudorapidity range.\cite{2010AtlasChargedParticles1,2010AtlasChargedParticles2}  CMS has published results  for the charged particle density of  non-single diffractive and inelastic events.
\cite{2010CmsChargedParticles1}$^-$\cite{2010CmsChargedParticles3}

\subsection{Charged particle correlations in high multiplicity events} 
\begin{figure}[pht]
\centerline{\psfig{file=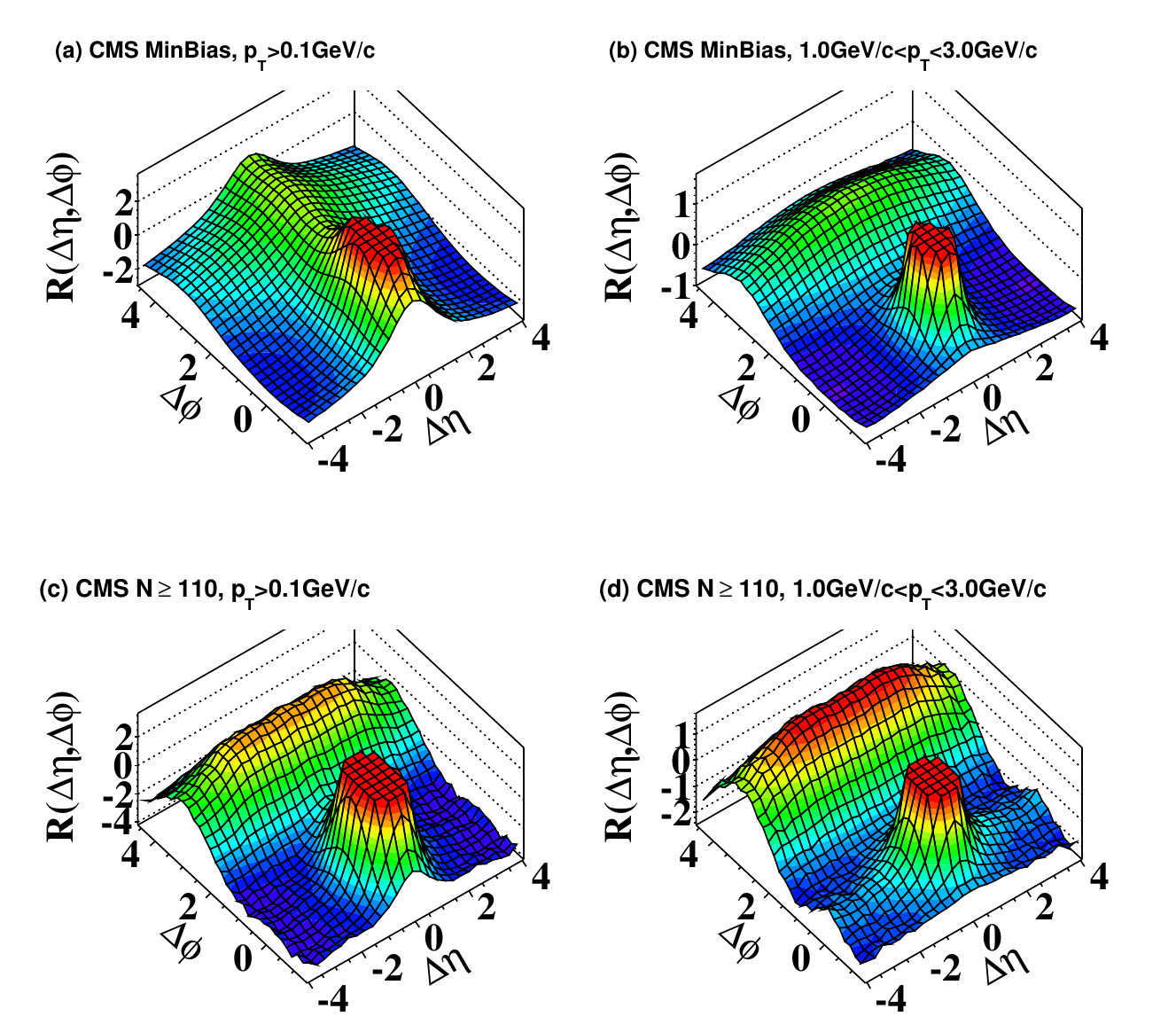,width=10cm}}
\vspace*{8pt}
\caption{(a),(b) CMS measurement of 2D two particle correlations for minimum bias events, and (c), (d) high multiplicity events ($n_{track}>110$).  The sharp near-side peak from jet correlations is cut off in order to better illustrate the structure outside that region. \protect\label{CMSCorrelations}}
\end{figure}
The CMS collaboration has observed an interesting effect in particle correlations in the two dimensional $\Delta\eta$ and  $\Delta\phi$ plane ($\phi$ is the azimuthal angle around the beam direction) as shown in Fig.~\ref{CMSCorrelations}.\cite{2010CMSCorrelations}   The observed structure in the correlation function at $\Delta\eta \approx  0$ and $\Delta\phi \approx  0$ can be interpreted as the onset of jet production and Bose-Einstein correlations. The backward rim at $\Delta\phi\approx \pi$ is the result of momentum conservation. The new observation of CMS is the appearance of a rim (long range correlation) at   $\Delta\phi \approx 0$ in high multiplicity events with the number of charged particles larger than 110 and $1~GeV< p_\perp<3~GeV$ 
(Fig. \ref{CMSCorrelations}(d)). A similar effect had been observed previously at RHIC in heavy ion collisions.\cite{RhicHiCorrelation} This is the first time that the effect is also seen in proton-proton collisions. The physical interpretation of this rim is subject to further investigations.

\subsection{Bottom quark production}
The LHC is a factory of $b$-flavored hadrons. Compared to $b\bar{b}$ production in $e^+ e^-$ collisions, one expects a higher production cross-section, but a larger background. The main goal at the LHC is to measure CP violation and rare decays of hadrons containing $b$- and $c$-quarks. The LHCb detector is dedicated for the study of $c$- and $b$-quarks. Its detector elements are placed along the beam line of the LHC, starting with a vertex locator, followed by trackers, ring imaging cherenkov detectors, calorimeters and a muon spectrometer. One of the first results of the LHCb collaboration was the measurement of the production cross-section of $b$-flavored hadrons.\cite{2010LHCbBxsec}       

\begin{figure}[pht]
\centerline{\psfig{file=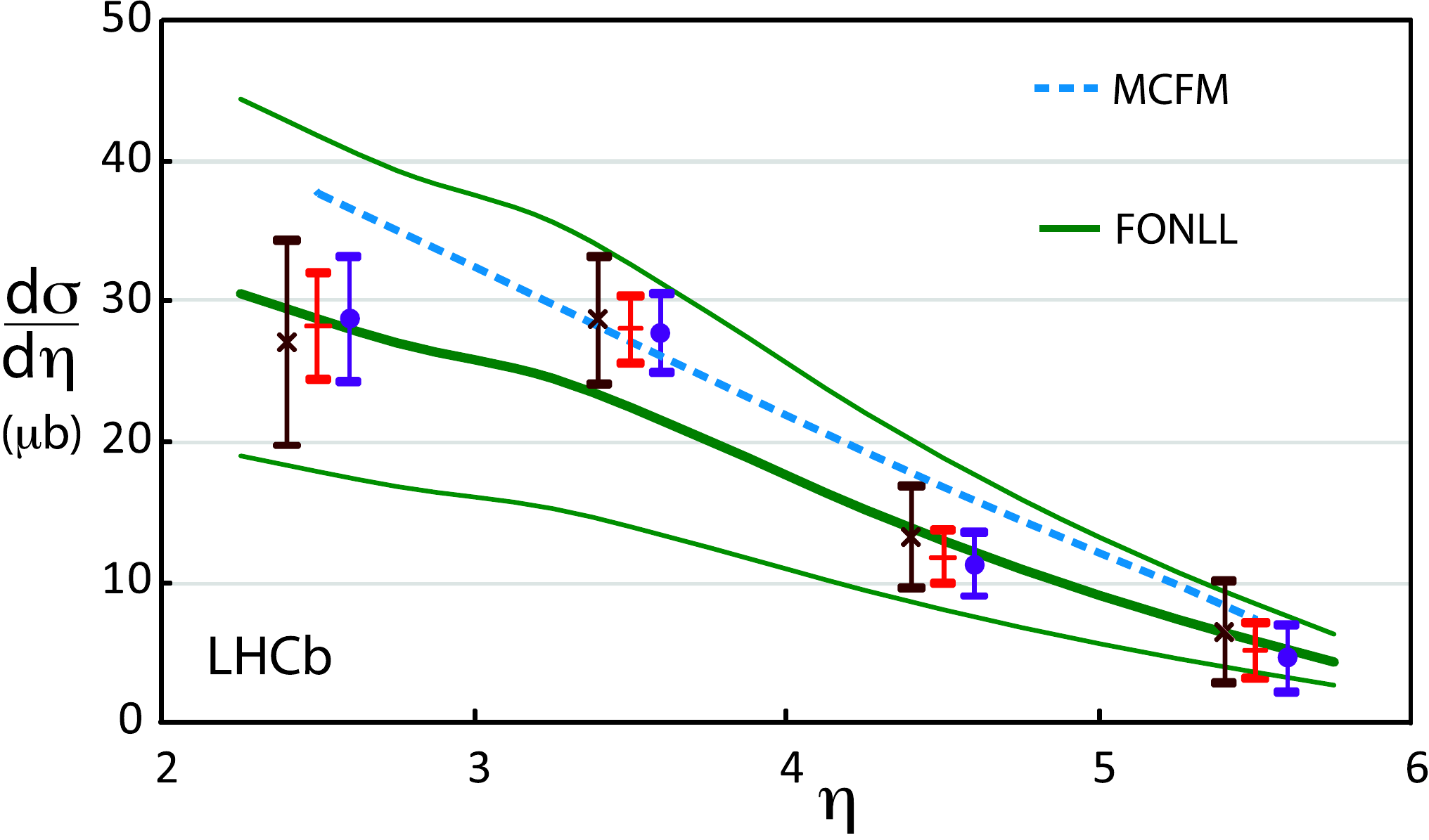,width=10cm}}
\vspace*{8pt}
\caption{LHCb measurement of $\sigma(pp \rightarrow H_bX$) at $\sqrt{s}=7$~TeV as a function of $\eta$ for the microbias ($\times$) and muon triggered ($\bullet$) samples, shown displaced from the bin center and the average (+). The data are shown as points with error bars, the MCFM prediction as a dashed line, and the FONLL prediction as a thick solid line. The thin upper and lower lines indicate the theoretical uncertainties on the FONLL prediction. The systematic uncertainties in the data are not included. \protect\label{LHCbBxsec}}
\end{figure}

Figure~\ref{LHCbBxsec} shows the result of the measurement for two different triggers, one microbias trigger and a muon based trigger. The measurement is in good agreement with two theoretical calculations, MCFM\cite{MCFM} and FONLL.\cite{FONLL} An extrapolation of the measured cross-section to the full pseudorapidity range yields a total   $b\bar{b}$ cross-section of 
\begin{equation}
\sigma(pp\rightarrow b\bar{b}X)=(284\pm20\pm49)\ \mu b
\end{equation}
using the LEP fragmentation function for $b$-quarks.  

Other results of the $c$- and $b$-flavored hadron production have been published by the CMS collaboration, \cite{2010CmsBquarks1}$^-$\cite{2010CmsBquarks3} where the production cross-sections of $J/\psi$, Y and reconstructed B mesons are measured.  

LHCb has obtained a first result on the branching ratio $B_s \rightarrow \mu^+\mu^-$. The Standard Model predicts  a value of $(0.32 \pm 0.02) \times 10^{-8}$, but new particles, for example supersymmetric particles, could increase this branching ratio substantially. A precise measurement of the branching ratio is therefore one of the best tests for new physics in B meson decays.   From the 2010 data LHCb obtains the 95\% C.L. limits\cite{2011LHCbBsToMumu}
\begin{eqnarray}
\mathrm{Br}(B^0_s\rightarrow \mu^+\mu^-) &<& 5.6 \times 10^{-8},\\
\mathrm{Br}(B^0_d\rightarrow \mu^+\mu^-) &<& 1.5 \times 10^{-8}.
\end{eqnarray}
These limits, obtained with $37~\mathrm{pb}^{-1}$, are comparable to limits obtained at the Tevatron with about $6~\mathrm{fb}^{-1}$.  

\subsection{W and Z boson production}

\begin{figure}[pht]
\centerline{\psfig{file=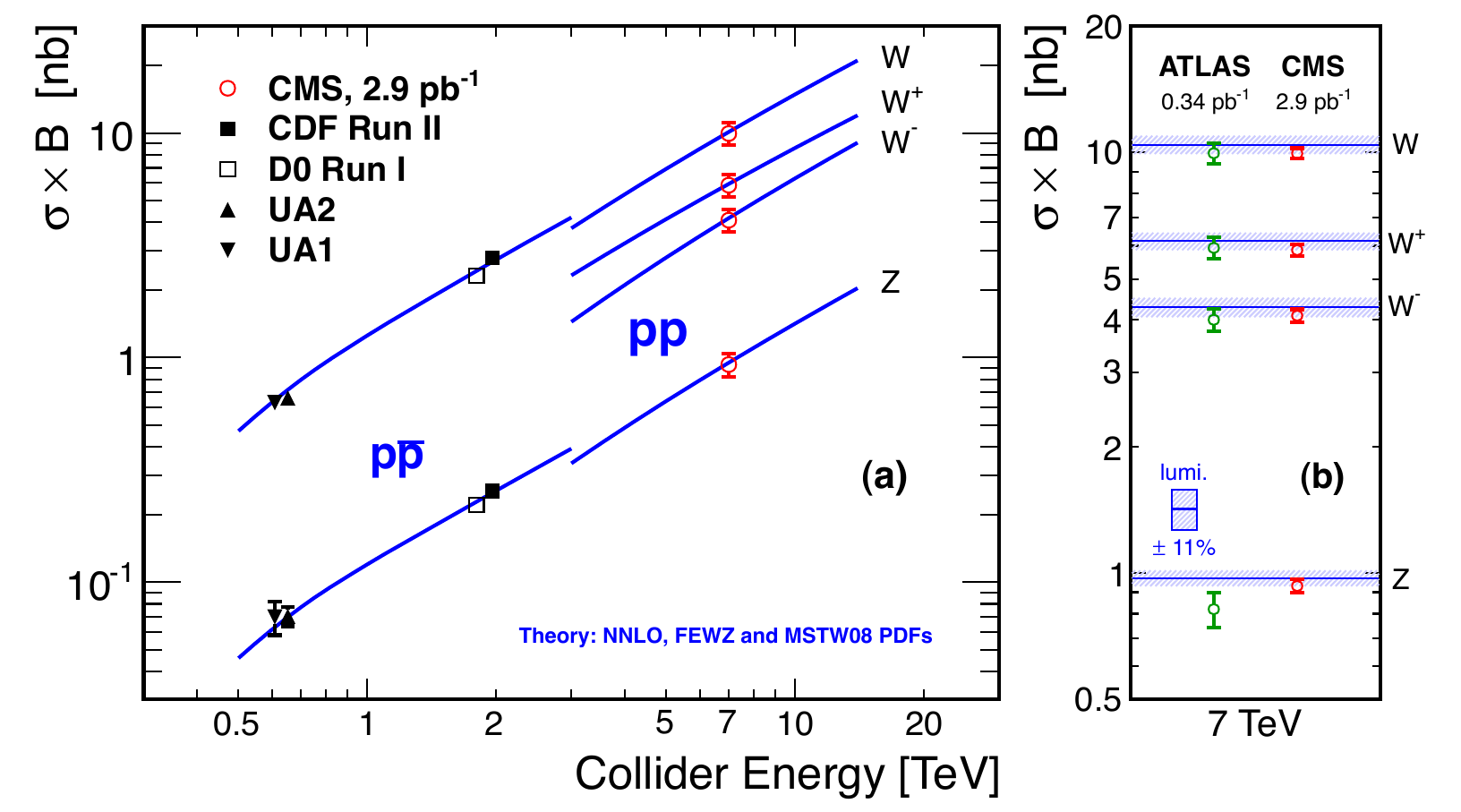,width=12cm}}
\vspace*{8pt}
\caption{(a) Measurements of inclusive W and Z production cross-sections times branching
ratio as a function of center-of-mass energy for CMS, ATLAS and experiments at lower-energy colliders.
The lines are the NNLO theory predictions. (b) Comparison of the ATLAS and CMS W and Z
production cross-sections times branching ratios. The error bars are the statistical and systematic
uncertainties added in quadrature, except for the uncertainty on the integrated luminosity,
whose value is shown separately as a band (Figure from Ref. \protect\refcite{2010CmsWz}). \protect\label{CmsWzRoots}}
\end{figure}

\begin{table}[h]
\tbl{Measurements of the inclusive W and Z cross-sections by ATLAS and CMS. The first error is statistical and the second systematic, an additional 11\% luminosity error applies to the cross-section measurements but not to the ratio $R_{WZ}$. }
{\begin{tabular}{@{}llll@{}} \toprule
Quantity & ATLAS($0.32~pb^{-1}$) & CMS($2.9~pb^{-1}$)&Theory(NNLO) \\
\colrule
$\sigma_{W\ell}=\sigma_W\cdot Br(W\rightarrow \ell\nu)$ 
& $9.95\pm0.07\pm0.28 $~nb
&$9.96\pm0.23\pm0.50 $~nb
&$10.44\pm0.52$~nb \\
$\sigma_{Z\ell\ell}=\sigma_Z\cdot Br(Z\rightarrow \ell\ell)$ 
& $0.82\pm0.06\pm0.05 $~nb 
& $0.93\pm0.02\pm0.10 $~nb 
&$0.972\pm 0.042$~nb\\
$R_{WZ}=\sigma_{W\ell}/\sigma_{Z\ell\ell}$
& $11.7\pm0.9\pm0.4 $ 
& $10.64\pm0.28\pm0.004 $ 
& $10.840\pm0.054$\\ \botrule
\end{tabular}\label{WZXsec} }
\end{table}
An important  milestone for the LHC experiments in 2010 was the measurement of the production cross-sections of W and Z bosons.  The backgrounds from QCD processes are small because one selects charged leptons (electron or muons) with high transverse momentum (typically $p_t > 20$~GeV) and missing transverse energy (for W) or a second oppositely charged lepton (for Z).  Table~\ref{WZXsec} shows the measured cross-sections for W and Z bosons by ATLAS\cite{2010AtlasWz} and CMS.\cite{2010CmsWz} The first error is statistical and the second systematic. In addition, a luminosity error of 11\% is assigned to the cross-section measurements.  An important test of the electroweak theory is the ratio of the W and Z production cross-section, because many systematic errors like the luminosity cancel. All measurements show very good agreement with NNLO calculations as shown in Fig.~\ref{CmsWzRoots} where the LHC measurements are compared to results at lower energies. 

ATLAS and CMS have also measured the charge asymmetry of inclusive semi-leptonic W decays as function of the pseudorapidity of the lepton. This measurement provides information on the $u$- and $d$-quark momentum fractions in the proton.\cite{2011AtlasWasym,2011CmsWasym} 

\subsection{Top quark production}
\begin{figure}[pht]
\centerline{\psfig{file=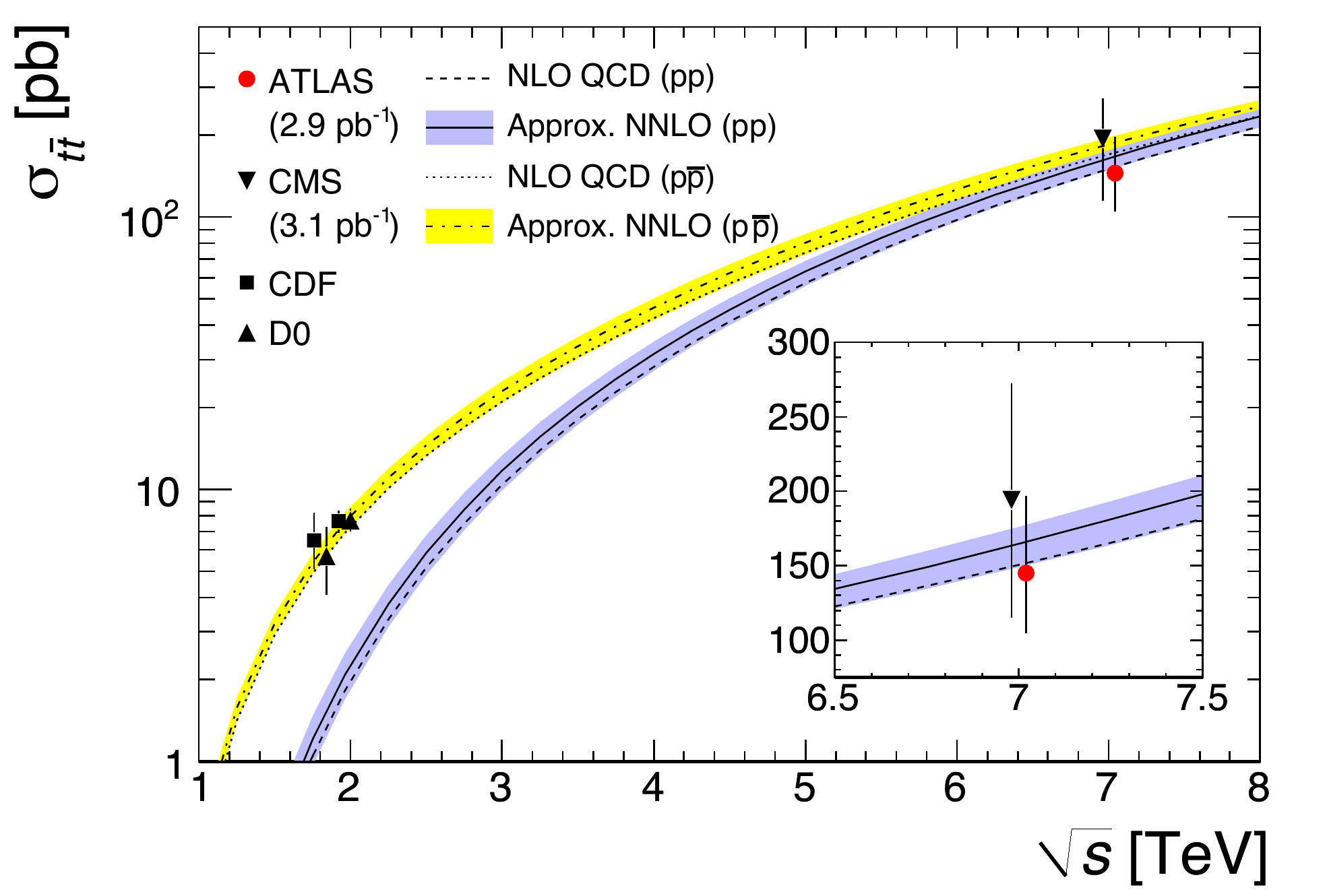,width=12cm}}
\vspace*{8pt}
\caption{Top quark pair-production cross-section at hadron colliders as measured by CDF and D0 at Tevatron, CMS and ATLAS. The theoretical predictions for $pp$ and $p\bar{p}$ collisions include the scale and PDF uncertainties. (Fig. from Ref. \protect\refcite{2010AtlasTopXsec}) \protect\label{AtlasTopXsec}}
\end{figure}

The  production cross-section for pairs of top quarks via gluon fusion increases steeply with center-of-mass energy. For $\sqrt{s}=7$~TeV one expects in the Standard Model (SM) a cross-section of $164^{+11.4}_{-15.7}$~pb  for $m_t=172.5$~GeV.\cite{TopXsec}  The production cross-section has been measured by CMS\cite{2010CmsTopXsec} and ATLAS\cite{2010AtlasTopXsec} using events with one lepton or two leptons plus jets.  The event selection is based on the observation of electrons or muons with missing energy and tagged $b$-quark jets. 

The ATLAS measurement uses an integrated luminosity of $2.9~\mathrm{pb}^{-1}$. 37 $t\bar{t}$ candidates are observed in the single-lepton topology, and nine candidates in the two-lepton topology, resulting in a $t\bar{t}$ cross-section of 
\begin{equation}
\sigma_{t\bar{t}} =145\pm31^{+42}_{-27}~\mathrm{pb}. 
\end{equation}
The measurement is in good agreement with the CMS measurement, based on the di-lepton topology, as shown in Fig.~\ref{AtlasTopXsec}. Further references to the Tevatron measurements and to the  theoretical calculations can be found in Ref. \refcite{2010AtlasTopXsec}. All the $t\bar{t}$ cross-section measurements at hadron colliders agree with the Standard Model predictions. 

\subsection{Search for New Particles}

The main motivation to construct the LHC was to search for the Higgs boson and for new effects which point to physics beyond the Standard Model, for example supersymmetry  or extra-dimensions. With the expected integrated luminosity of 1~fb$^{-1}$ or more at the end of 2011 it should be possible to constrain such models severely or observe new effects and particles. In 2010, only a luminosity of 35~pb$^{-1}$ per experiment has been analysed. Already with this integrated luminosity ATLAS and CMS have published many limits on the production of  new particles, for example excited quarks, lepto-quarks, extra dimensions, microscopic black holes and $W^\prime$ bosons, exceeding significantly the limits of Tevatron, LEP and HERA.   So far no effect has been observed which points to deviations from the Standard Model. The present limits and the satisfactory control and understanding of backgrounds lead to promising perspectives on particle searches for the 2011 data analysis. 

\begin{center}
\begin{figure}[pht]
\psfig{file=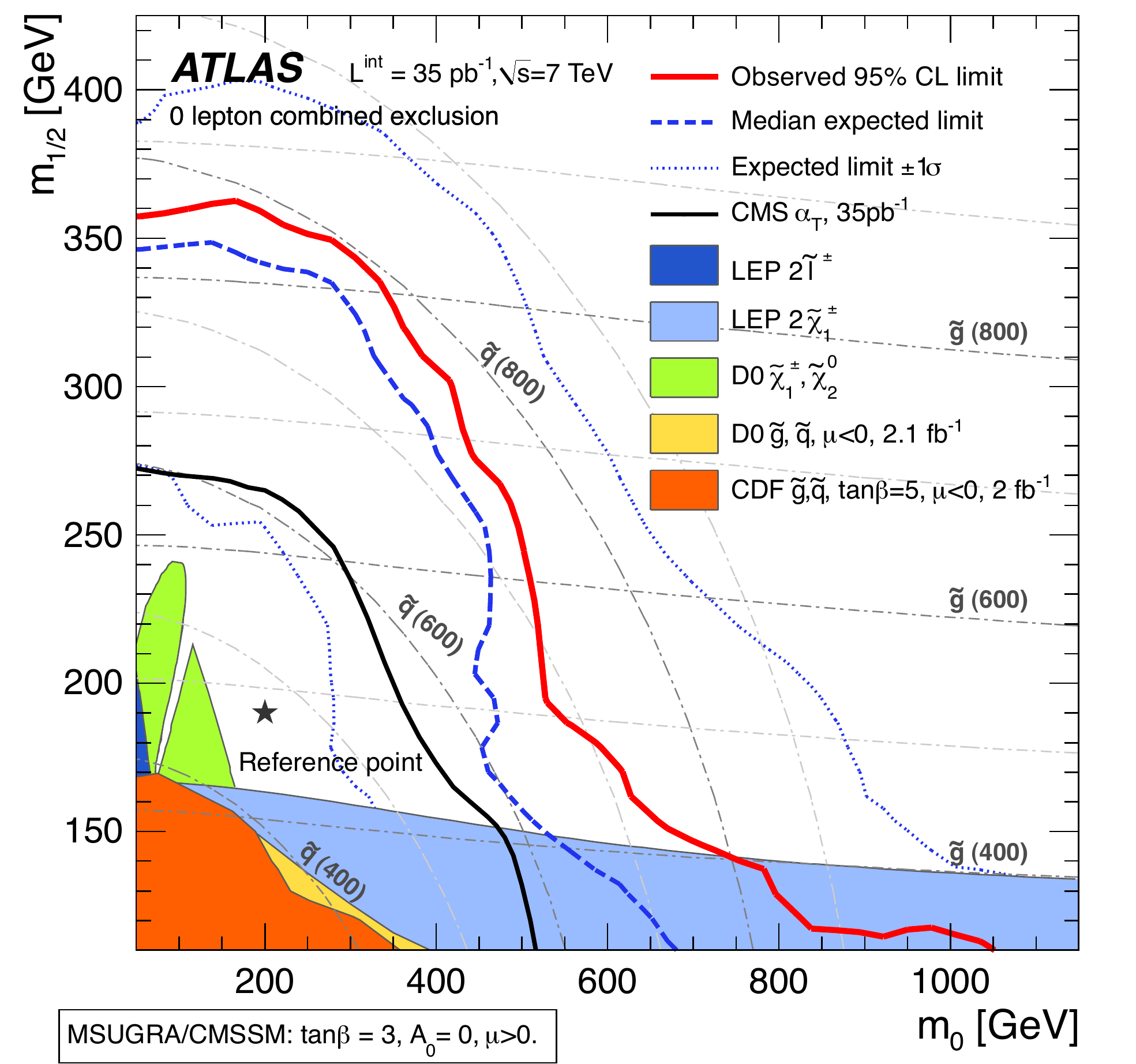,width=7.5cm}
\psfig{file=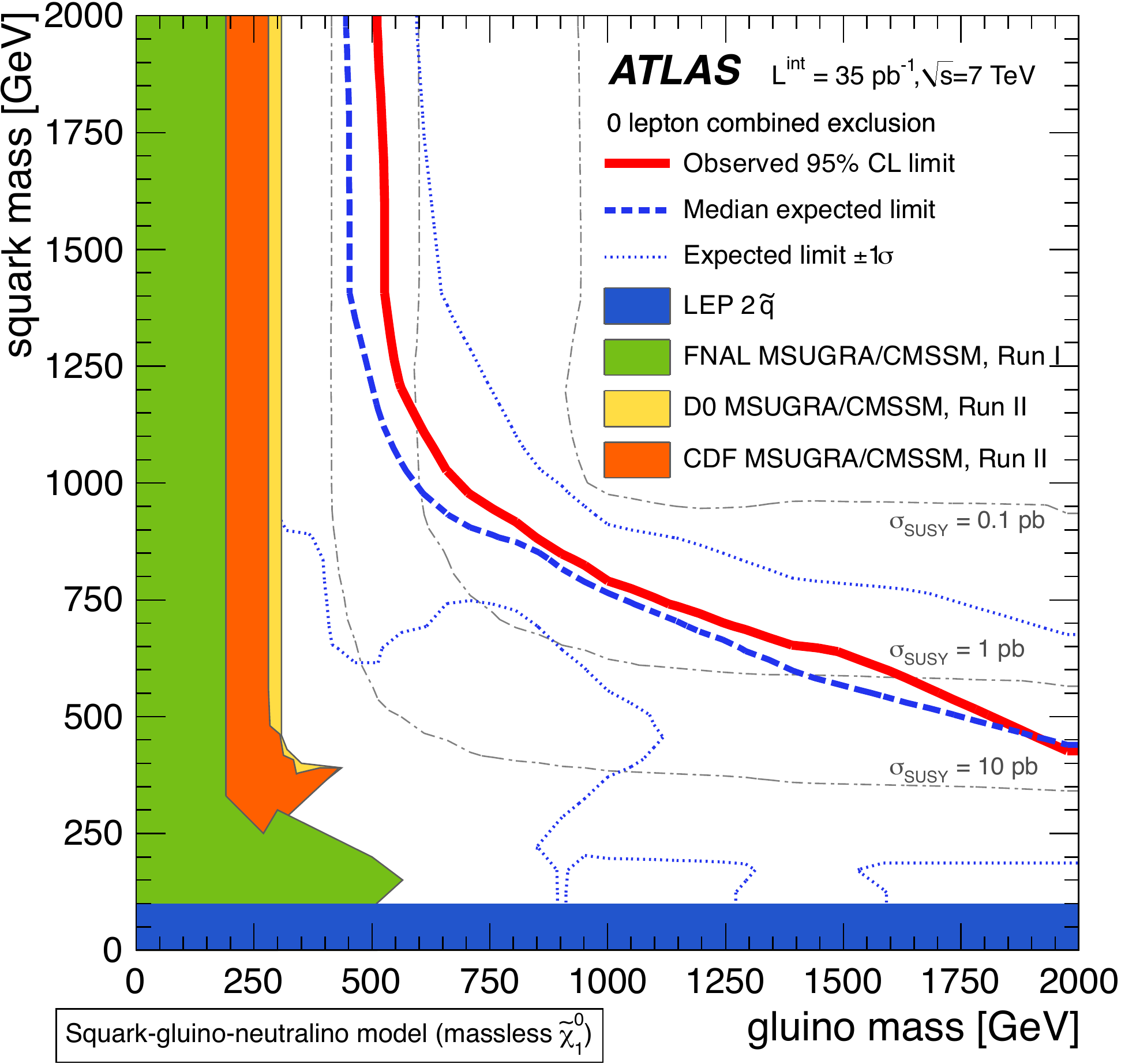,width=7.5cm}

\vspace*{8pt}
\caption{Left: 95\% C.L. exclusion limits in the $\tan\beta=3,\ A_0=0$ and $\mu>0$ slice of MSUGRA/CMSSM, together with existing limits with the different model assumptions given in the legend. 
Right: 95\% C.L. exclusion limits in the (gluino, squark) mass plane together with existing limits. 
Comparison with existing limits is illustrative only as some are derived in the context of MSUGRA/CMSSM or may not assume a massless neutralino.   (Fig. from Ref. \protect\refcite{2011AtlasSusy0Lepton}) \protect\label{AtlasSusy0Lepton}
}
\end{figure}\end{center}

One of the theoretically favored theories beyond the Standard Model is supersymmetry, which predicts partners of standard model particles which differ in spin by 1/2. At the LHC the cross-sections of strongly interacting squarks and gluinos, the partners of quarks and gluons, are expected to be large. This has enabled ATLAS and CMS to  performed the first searches for supersymmetric particles, based on events with many jets, missing transverse momentum and with or without leptons and photons.\cite{2011CmsSusy0Lepton}$^-$\cite{2011AtlasSusyBJet}
No excess with respect to the Standard Model predictions has been observed. These results allow to set 95\% C.L. limits on the parameter space of SUSY models. Figure \ref{AtlasSusy0Lepton} shows the  95\% C.L. exclusion limits in the $M_0-M_{1/2}$ plane of the CMSSM/MSUGRA model obtained from an analysis of events with many jets and missing transverse momentum.  ATLAS and CMS are able to extend the previously excluded region significantly, for example in the specific case 
$m_{\tilde{q}} = m_{\tilde{g}}$ squark and gluino masses below 775~GeV are excluded in the CMSSM/MSUGRA model.

\subsection{Heavy Ion Physics}
At the end of the 2010 proton-proton run, lead ions were collided in the LHC at a nucleon-nucleon center-of-mass energy of $\sqrt{s_{NN}}=2.76$~TeV.  This data was eagerly awaited by the heavy-ion community, because it represents a significant boost in energy compared to  $\sqrt{s_{NN}}=0.2$~TeV at RHIC and will provide severe constrains for models of nucleus-nucleus collisions and on the formation of a quark-gluon plasma. The ALICE experiment at LHC is specialized on the investigation of heavy-ion collisions, but also ATLAS and CMS, due to their excellent tracking and calorimetry, are expected to provide important complementary results.      

\begin{figure}[pht]
\centerline{\psfig{file=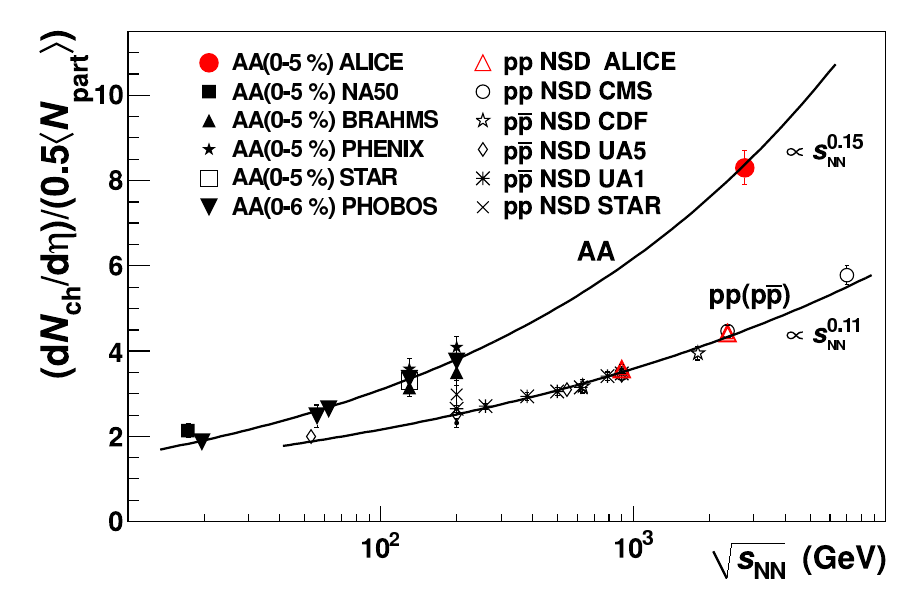,width=10cm}}
\vspace*{8pt}
\caption{Charged particle pseudorapidity density per participant pair for central-nucleus, and non-single diffractive $pp/p\bar{p}$ collisions, as a function of $\sigma_{NN}$ (Figure from Ref. \protect\refcite{2010AliceHiChargedParticles}) \protect\label{AliceHiChargedParticles}}
\end{figure}
For about one month lead-lead data was taken, corresponding to an integrated luminosity of about $10~\mu b^{-1}$. Within a few weeks several crucial measurements were published by the experiments. 
ALICE measured a multiplicity density of primary charged particles at mid-rapidity of
$1584 \pm 4$ (stat.)$ \pm 76$ (sys.) for the most central (5\%) collisions, which corresponds to $8.3 \pm 0.4$ (sys.) per participating nucleon pair.\cite{2010AliceHiChargedParticles} This represents an increase of about a factor 1.9  relative to pp collisions at similar collision energies, and about a factor 2.2 to central AuÐAu collisions at RHIC as shown in Fig.~\ref{AliceHiChargedParticles}. This first measurement already constrains and rules out many models of nucleus-nucleus collisions at LHC energies. 

ALICE\cite{2010AliceHiEllipticalFlow} also measured the charged particle elliptical flow in the central rapidity region $\vert\eta\vert<0.8$ as predicted by hydrodynamical models of heavy-ion collisions and observes an increase of 30\% compared to Au-Au collisions at RHIC. This increase is higher than predicted from ideal  hydrodynamic models, but agrees with models which contain viscous corrections. 

ATLAS has published two interesting observations, the suppression of jet-formation ("jet quenching")\cite{2010AtlasJetQuenching} and the suppression of $J/\psi$ production\cite{2010AtlasJpsiSuppression} in central lead-lead collisions. Both effects are predicted to be consequences of the formation of a quark-gluon plasma. 

\begin{figure}[pht]
\centerline{\psfig{file=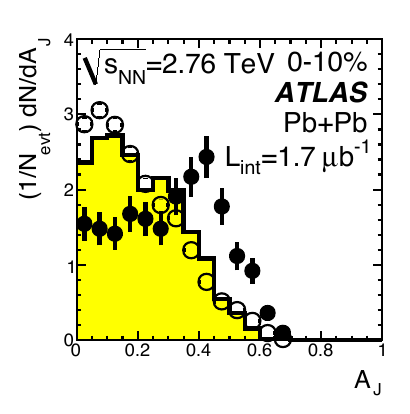,width=6.5cm}
\psfig{file=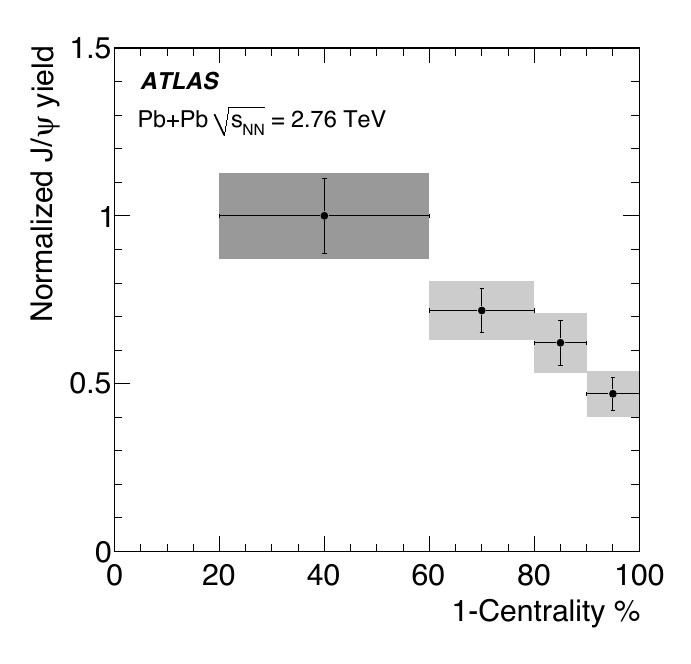,width=6.5cm}}
\vspace*{8pt}
\caption{Left: Measurement of the asymmetry in the transverse energy of two jets for the most central collisions.   Right: Normalized $J/\psi$ yield divided by the number of binary collisions as function of centrality (Figures from Refs. \protect\refcite{2010AtlasJetQuenching} and \protect\refcite{2010AtlasJpsiSuppression}). \protect\label{Atlas-JetQuenching}}
\end{figure} 

A measure for jet-quenching is the asymmetry in transverse energies of two jets:
\begin{equation}
A_j =\frac{E_{T1}-E_{T2}}{E_{T1}+E_{T2}}, \; \Delta \phi > \frac{\pi}{2}.
\end{equation}
A large asymmetry indicates that one jet in a hard scattering process escaped the interaction region unhindered, but the second lost a lot of transverse energy in the interaction with quarks and gluons. That could be the case   for collisions which happen at the boundary of the dense nucleus-nucleus interaction region. ATLAS has measured $A_J$ as function of centrality in lead-lead collisions. It finds, as shown in Fig.\ref{Atlas-JetQuenching}~(left),  that for central collisions the asymmetry is larger than expected in models like Hijing, which do not include jet-quenching effects, indicating that a very dense medium is formed in Pb-Pb collisions.     
This observation is supported by an ALICE publication\cite{2010AliceHiParticleQuenching} which shows a strong suppression of high $p_t$ charged particles in central collisions and by a CMS publication which provides more detailed studies of the energy distributions in events with jets.\cite{2011CmsHiJetQuenching} 

The second observation of ATLAS,\cite{2010AtlasJpsiSuppression} as shown in Fig.~\ref{Atlas-JetQuenching}~(right),  
demonstrates that $J/\psi$ production normalized to the number of binary collisions is strongly suppressed for central compared to peripheral Pb-Pb collisions in agreement with predictions from the formation of a dense media with colored constituents. 

The first heavy ion run at LHC has already provided a significant advancement in the understanding of Pb-Pb collisions with important constraints on theoretical models. A new door has been opened to a sequence of detailed studies in the coming years. 

\section{Summary}

The first year of LHC data taking provided an integrated luminosity of about $35~\mathrm{pb}^{-1}$ in proton-proton collisions at $\sqrt{s}=7$~TeV for physics analysis. Already in the first months the LHC experiments reached the design performance. In more than hundred publications and conference notes tests of the Standard Model and searches for new particles were performed.  Among other results the physics highlights have been the measurements of the W-, Z-boson and $t\bar{t}$ production cross-section, improved limits on supersymmetric and other hypothetical particles and the observation  of jet-quenching, elliptical flow and $J/\psi$ suppression in lead-lead collisions 
at $\sqrt{s_{_{NN}}} = 2.76$~TeV.  For the 2011 data taking one expects more than 50 times higher integrated luminosities. This will open the possibilities to search for the Higgs boson and to explore a wide parameter space in supersymmetric models. With the recent decision to continue the LHC data taking in 2012 at 7 or 8~TeV it should be possible to place 95 \% C.L. limits on Standard Model Higgs bosons from 114 GeV up to 600 GeV at the end of 2012. Therefore the coming two years will be decisive for the validity of the Standard Model and for the search for new particles.

\section*{Acknowledgments}

The author thanks the LHC machine staff for the successful operation of the accelerator and the colleagues from the experiments for providing such an extraordinary rich spectrum of physics results for this review article.

\end{document}